\documentclass[twocolumn,preprintnumbers,floatfix,prb,showpacs]{revtex4}
\usepackage{graphicx}
\usepackage{dcolumn} 
\usepackage{bm}
\usepackage{epsfig}
\usepackage{longtable}
\usepackage{multirow}
\usepackage{afterpage}
\usepackage{amsmath,amssymb}
\begin{document}
\title{Disturbing the dimers: electron- and hole-doping\\
   in the intermetallic insulator FeGa$_3$}
\author{Antia S. Botana}   
\email[Corresponding author. Email address: ]{abotana@anl.gov}
\affiliation{University of California-Davis, Davis, California, USA}
\author{Yundi Quan}
\affiliation{University of California-Davis, Davis, California, USA}
\author{Warren E. Pickett}
\email[Corresponding author. Email address: ]{wepickett@ucdavis.edu }
\affiliation{University of California-Davis, Davis, California, USA}
\pacs{71.20.Lp, 75.10.Lp,71.20.-b}
\date{\today}
\begin{abstract}
Insulating FeGa$_3$ poses peculiar
puzzles beyond the occurrence of an electronic gap in an intermetallic compound.  This Fe-based material has a very distinctive structural characteristic with the Fe atoms occurring in dimers.
The insulating gap can be described comparably well in either the
weakly correlated limit or the strongly correlated limit within density functional theory viewpoints,
where the latter corresponds to  singlet formation on the Fe$_2$ dimers. 
Though most of the calculated occupied Wannier functions are
 an admixture of  Fe $3d$
 and Ga $4s$ or $4p$ states, there is a single bonding-type Wannier function per spin centered on
 each Fe$_2$ dimer. 
 Density functional theory methods have been applied to follow the evolution
 of the magnetic properties and electronic spectrum with doping, where unusual
 behavior is observed experimentally.
 Both electron and hole doping are considered, by Ge and Zn on the
 Ga site, and by Co and Mn on the Fe site, the latter introducing direct
 disturbance of the Fe$_2$ dimer. Results from weakly and strongly correlated pictures
 are compared.
 Regardless of the method, magnetism including itinerant phases appears readily with doping.
 The correlated picture suggests that in the low doping limit
 Mn (for Fe) produces an in-gap hole state, while Co (for Fe) introduces a
 localized electronic gap state.
\end{abstract}
\maketitle

\date{\today}
\section{background}
FeGa$_3$ is a rare intermetallic insulator that has attracted particular attention due to its unusual transport 
and magnetic properties. This Fe-based material exhibits semiconducting behavior with a gap of 0.5 eV 
obtained from transport measurements.\cite{gap_1} Particular attention has been paid to understand 
the mechanism of the gap formation 
in the context of strong hybridization between Fe-$3d$ and Ga-$4p$ orbitals, reminiscent of that in 
strongly correlated $3d$ and $4f$ Kondo insulators.\cite{jphysconfser200_012014_2010} 

Given the small band gap and the presence of very narrow bands around the Fermi level, FeGa$_3$ has been most 
extensively studied as a thermoelectric material. High values of the Seebeck coefficient around 350 $\mu$V/K have been 
measured at room temperature.\cite{therm_exp} In single crystals colossal values of the thermopower ($\sim$ 
-16000 $\mu$V/K) emerge below 20 K due to the phonon drag effect.\cite{deepa_fega3}

There is no unambiguous picture about the role of electronic correlations and the magnetic (or not) character of stoichiometric FeGa$_3$ from experiments. According 
to susceptibility measurements, FeGa$_3$ is diamagnetic below room temperature (RT) and the susceptibility shows an 
increase above RT suggesting proximity to a crossover to a paramagnetic metallic state.\cite{gap_1} Fe M\"ossbauer 
spectra did not show the presence of an internal magnetic field at the Fe site, supporting a nonmagnetic state of Fe
\cite{mossbauer} though not ruling out correlated states. 
In contrast, muon spin rotation studies detected 
spectroscopic features interpreted in terms of electron confinement into spin polarons that require the existence of Fe 
moments.\cite{muon} The narrow gap in FeGa$_3$ and its unusual  properties suggesting electronic correlations brings to mind related iron compounds FeSi\cite{fesi_1, fesi_2,fesi_3} and  FeSb$_2$,\cite{fesb2_1, fesb2_2} whose underlying electronic systems remain to be fully understood.

Hole or electron doping drastically changes the properties of the parent compound, giving rise to emergent magnetic 
phases. FeGa$_3$ has a very distinctive structural characteristic: Fe atoms occur in dimers as shown in Fig. \ref{struct}. Replacing Fe breaks the dimer symmetry, a fundamental impact {\it if} there is important Fe-Fe bonding,
spin correlations, or singlet 
formation. Focusing first on electron doping, Co-substitution for Fe induces an insulator-to-metal transition.
\cite{jphysconfser200_012014_2010,PhysRevB.89.104426,PhysRevB.86.144421} Resistivity $\rho(T)$ measurements for 
Fe$_{1-x}$Co$_x$Ga$_3$ are not yet conclusive; some indicate the metallic state is reached at doping level $x
$=0.025-0.075,\cite{PhysRevB.89.104426}  in other reports only at substantially higher doping levels of $x
$=0.125\cite{jphysconfser200_012014_2010} or 0.23.\cite{PhysRevB.86.144421} 

Analysis of the T-dependence of the nuclear spin-lattice relaxation rate 1/T$_1$ of the $^{69,71}$Ga nuclei suggests the 
existence of in-gap states at low Co doping.\cite{PhysRevB.89.104426} In Fe$_{0.5}$Co$_{0.5}$Ga$_3$, the relaxation is 
strongly enhanced due to spin fluctuations, often a signature of a weakly antiferromagnetic (AFM) metal. Such itinerant 
antiferromagnetic behavior contrasts with magnetization measurements, showing localized magnetism with a 
relatively low effective moment of 0.7$\mu_B$/f.u.\cite{PhysRevB.89.104426}  

Electron-doping by substituting Ga by Ge leads to drastically different behavior. \cite{PhysRevB.86.144421, arxiv_ge} Experiments show that FeGa$_{3-y}$Ge$_y$ is conducting at an extremely low doping level $y$= 0.0006, progressing to a weak ferromagnetic order at $y$$_c
$= 0.13 which never appears in the Co-doped compound. The emergence of the ferromagnetic (FM) state is 
accompanied by quantum critical behavior observed in the specific heat and the magnetic susceptibility. The FM instability 
found in FeGa$_{3-y}$Ge$_y$ beyond $y$$_c$= 0.13 indicates that strong electron correlations are induced by the disturbance of Fe $3d$ -- Ga $4p$ hybridization, or possibly that existing strong correlations are disrupted. \cite{PhysRevB.86.144421} 

Turning now to hole doping, Gamza \textit{et al.}\cite{PhysRevB.89.195102} performed resistivity and thermodynamic 
measurements on single crystals of FeGa$_3$, Fe$_{1-x}$Mn$_x$Ga$_3$ and FeGa$_{3-y}$Zn$_y$ (x $\leq$ 0.12 and y 
$\leq$ 0.06). Unlike for electron doping, hole doping using Mn on the Fe site or Zn on the Ga site does not give rise to a 
semiconductor-to-metal transition. Hole doping induces states into the semiconducting gap that remain localized at the 
highest doping levels. Using neutron powder diffraction measurements, they conclude that FeGa$_3$ orders magnetically 
above room temperature in a complex structure, unaffected by the inclusion of Mn and Zn. 
 
Evidently input from theoretical modeling is required to move toward understanding of this unusual behavior. One emphasis of this paper is to give special attention to the metal dimers in this structure. The flat
bands bordering the gap suggest correlated electron behavior, and interatomic correlations should be much
stronger between electrons or spins within a dimer than between dimers, and local physics may appear. The 
situation can be illustrated by considering Fe$_{0.5}$Co$_{0.5}$Ga$_3$ mentioned above. Any given dimer has a
25\% chance of being Fe$_2$, 25\% chance of being Co$_2$, and 50\% chance of FeCo. These three configurations 
may have very different physics, and their states may lie in the gap (and may destroy it) or in the sea of
itinerant states. The system average may be very complex. Doping on the Ga site, however, does not directly
disturb the dimers, and the change in band filling may be easier to handle and to understand.  

DFT-based calculations for FeGa$_3$ have been reported  using the two more common
(semi)local approaches for the exchange correlation energy and potential: local density 
approximation (LDA)\cite{lda} and
generalized gradient approximation (GGA).\cite{gga} The very similar results 
seem to be sufficient for describing the electronic
structure of the undoped compound, giving a value\cite{PhysRevB.82.155202} 
of the gap of 0.4-0.5 eV comparable to the experiments.
However, those approaches are unable to model the AFM phase in Co-doped FeGa$_3$, and
they incorrectly predict a metallic FM ground state for hole doped FeGa$_3$, whether by Ga or Fe substitution.
Within the correlated DFT LDA+$U$ 
method (see Sec. VI), local moments arise on the Fe dimer if the spins are antialigned, giving an
antiferromagnetic result in band theory. 

In view of these results, Yin and 
Pickett\cite{PhysRevB.82.155202} suggested that the Fe dimers could be forming spin singlets -- strongly
correlated but nonmagnetic local states -- and that 
magnetism found in doped FeGa$_3$ would be linked to the breaking of the singlets into free spins. Singh showed
\cite{PhysRevB.88.064422} that the magnetism of some types of doping of FeGa$_3$ can be explained within GGA 
without the need of spin coupling of pre-existing moments. 
This weakly correlated picture suggests that both $n$-type and 
$p$-type FeGa$_3$ will become itinerant ferromagnets\cite{PhysRevB.88.064422} due to the large density of states on 
either side of the gap. Recently, it has been proposed from DFT-based calculations combined with 
Fe Moessbauer spectroscopy that increasing Ge-doping level ($y$) in FeGa$_{3-y}$Ge$_y$  over a 
wide range 0.03$\leq$$y$$\leq$0.5 
leads to an evolution from localized moments to a combination of localized and itinerant moments with 
the interplay of ferromagnetism and antiferromagnetism.\cite{arxiv_gedoped} 
Dynamical mean field theory (DMFT) calculations have also been performed in 
FeGa$_3$,\cite{PhysRevB.89.195102} showing that Fe ions 
are dominantly in an $S=$1 state displaying strong spin and charge fluctuations. The DMFT approach is
unable to model interatomic spin correlations.

The use of a non-multiplicative (non-local) potential such as LDA+$U$ and the explanation in terms of an 
antiferromagnetic ground state for undoped FeGa$_3$\cite{PhysRevB.82.155202} introduces new questions about the 
role of strong electronic correlations. In this paper 
we revisit several aspects of the electronic structure of electron and hole-doped FeGa$_3$, comparing results from the 
LDA and  LDA+$U$ functional forms. Because in several cases the value of the gap is of some importance, in the 
appendix we indicate how the modified Becke-Johnson potential (see within) shifts bands and in some cases modifies 
magnetic moments giving better agreement with the experimental results.

\section{structure}

\begin{figure}[!ht]
\includegraphics[width=0.95\columnwidth,draft=false]{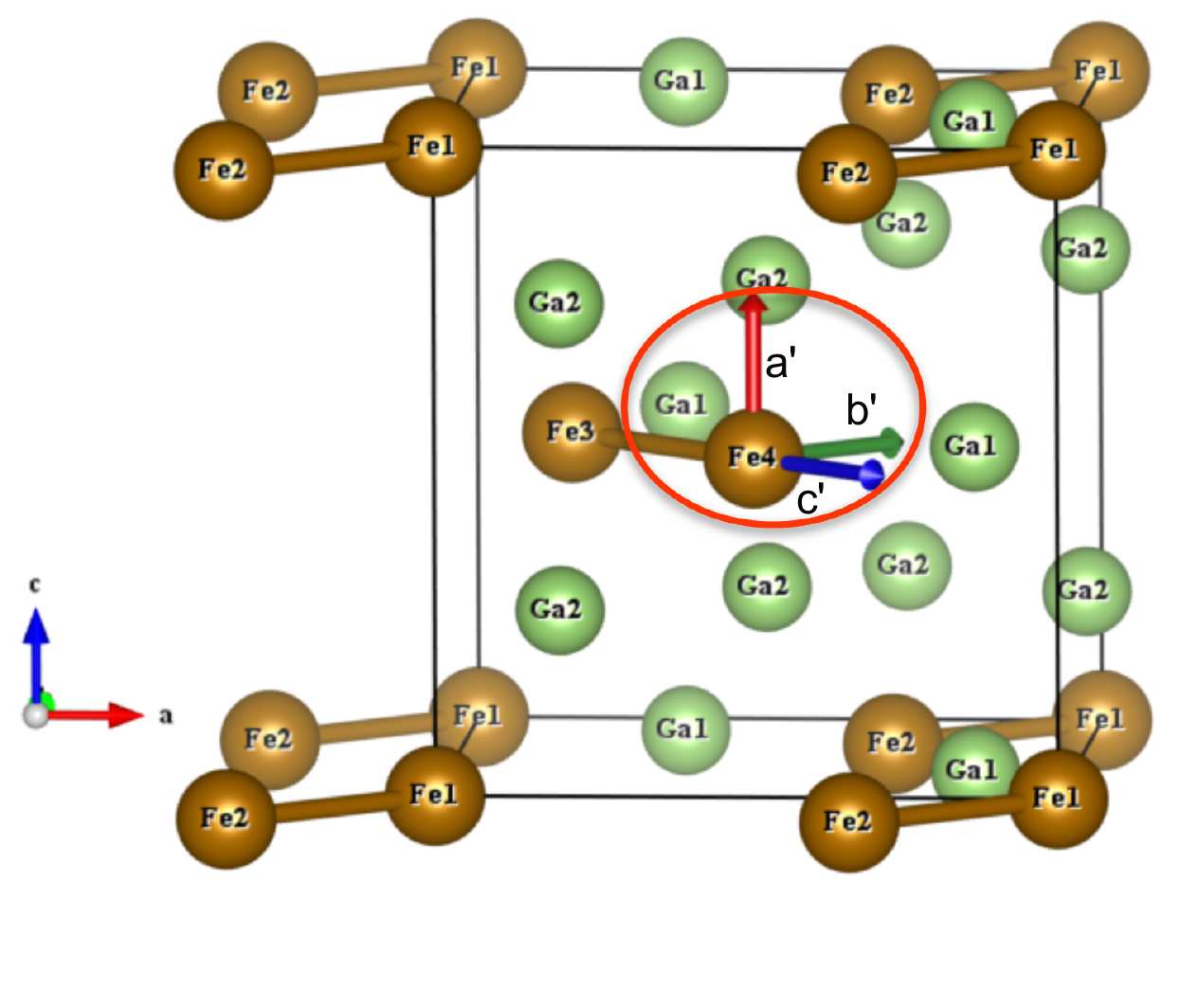}
\includegraphics[width=0.85\columnwidth,draft=false]{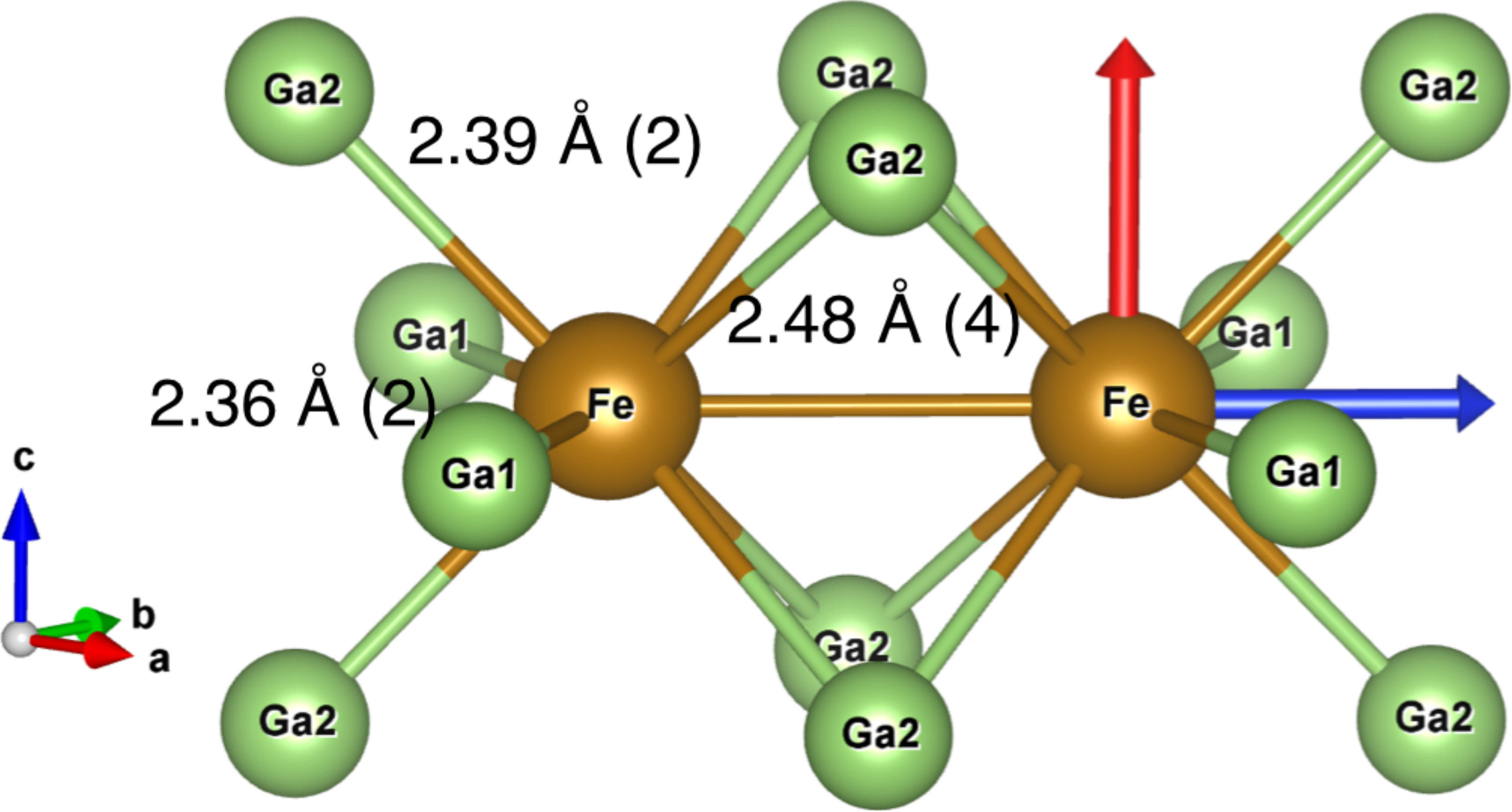}
\caption{\label{fignew}(Color online) Upper panel: Crystal structure of the unit cell of FeGa$_3$. Fe atoms form dimers 
oriented along the [110] and [1$\bar{1}$0] directions. The local coordinate system used in the DOS plots with the local $z$ 
axis directed along the Fe$_2$ dimers is circled. Lower panel:  Local environment of Fe 
atoms in a dimer formed by Ga2 at 2.39(2) and 2.48(4) \AA ~and Ga1 at 2.36(2) \AA. The two distinct sites for Ga atoms 
can be clearly identified. The local coordinate system used in the density of states plots is also shown. The unit cell shown corresponds to a structure with space group \textit{Pm} (lower symmetry than the experimental one, see Section III).}
\label{struct}
\end{figure}

FeGa$_3$ crystallizes in the tetragonal space group P4$_2$/mnm. Its lattice constants are $a$=
6.2628 \AA ~and $c$= 6.5546 \AA ~with four formula units (two Fe$_2$ dimers) per unit cell.  Fe atoms are at (0.3437, 
0.3437, 0) and form dimer pairs in the $z$=0 plane along (110) and in $z$= 1/2 along the $(1\bar{1}0)$ directions, as 
illustrated in Fig. \ref{struct}. There are two inequivalent Ga sites: higher symmetry Ga1 at (0, 0.5, 0), and lower symmetry 
Ga2 at (0.1556, 0.1556, 0.262). Each Fe atom has eight Ga neighbors, two Ga1 at distances of 2.36 \AA, and six Ga2, two 
of them at 2.39 and four at 2.48 \AA ~(see Fig. \ref{struct}). The paired Fe atoms are separated by 2.77 \AA, 12 $\%$ 
larger distance than that between Fe atoms in bcc Fe metal (2.48 \AA).

\section{Computational Methods}\label{methods}

The electronic structure calculations were performed with the WIEN2k code,\cite{wien2k,wien} based on density functional 
theory (DFT) utilizing the augmented plane wave plus local orbitals method (APW+lo).\cite{sjo}
The calculations were well converged
with respect to $k-$mesh and R$_{mt}$K$_{max}$, using up to 1000
k-points (10$\times$10$\times$9 mesh) and R$_{mt}$K$_{max}$=7.0 cutoff.
Selected sphere radii (a.u.) were
the following: 2.27 for Fe, 2.16 for Ga, 2.35 for Mn and Co, 2.16 for Ge, and 2.22 for Zn.

Because the description of the electronic structure is in question, an assortment of exchange-correlation potentials have 
been used: LDA,\cite{lda} LDA+$U$ (LDA plus the on-site repulsion $U$ within the fully localized limit functional),\cite{sic} and the Tran-Blaha modified Becke 
Johnson (from hereon mBJ for simplicity) included with LDA. 

The LDA+$U$ scheme improves over GGA or LDA in
the study of systems containing strong intraatomic repulsion such as occurs in many transition metal compounds.
\cite{sjo,sic}  We have calculated the effective $U$ for FeGa$_3$ using the approach proposed 
by Madsen and Novak\cite{madsen_u} for augmented plane wave methods based on 
the procedure of Anisimov and Gunnarsson.\cite{anisimov} In this 
method the occupation of a target orbital (in this case the 3$d$ orbitals of Fe) 
is enforced. Due to the ambiguity in the charge state of Fe (see discussion below), different 3$d$ occupations from $d^8$ to $d^5$ were presumed, with
resulting effective $U$ values increasing linearly from 2 to 5 eV, respectively. 
We have performed calculations using the fully localized limit double counting
functional within this $U$ range, with $J$ being set to 0.8 eV. The same value of $U$ has been applied 
simultaneously to Fe, and to neighboring atoms Mn and Co. 

The mBJ exchange potential (a local
approximation to an atomic exact-exchange potential and
a screening term) + LDA correlation
allows the calculation of band gaps with an
accuracy similar to the much more expensive GW or hybrid
methods. \cite{PhysRevLett.102.226401, PhysRevB.83.195134} We have also studied the electronic 
structure of both undoped and doped 
FeGa$_3$ by using the mBJ potential which does not
contain any system-dependent parameter.

Calculations with the magnetic moments within the metal dimers being aligned and antialigned have been performed. To be able to establish different magnetic orderings, a lower symmetry structure (with space group \textit{Pm}) was used. This structure contains 12 inequivalent atoms: four inequivalent Fe sites (Fe1-Fe2 forming one dimer and Fe3-Fe4 forming the 
other dimer), four inequivalent Ga1 sites, and four inequivalent Ga2 sites (see Fig. \ref{struct}).  

Table \ref{table1} shows, for each of the doping mechanisms and computational schemes studied, the magnetic moments of each of the atoms in the metal dimers, the total magnetic moment in the cell, and the band gap for the corresponding magnetic ground state.

\begin{table*}[!ht]
\caption{Fe/Mn/Co atomic moment (M, in $\mu_{B}$), total magnetic moment in the unit cell (M$_{tot}$, in $\mu_{B}$), and band gap (in eV) for the magnetic ground state of undoped and hole/electron doped FeGa$_3$ within LDA, LDA+$U$ ($U$= 3 eV, $J$= 0.8 eV), and mBJ.}
\begin{ruledtabular}
\begin{tabular}{lccccc}
\multicolumn{1}{l}{} &
\multicolumn{1}{c}{M Fe1/Fe2} &
\multicolumn{1}{c}{M Fe3/Fe4} &
\multicolumn{1}{c}{M$_{tot}$} &
\multicolumn{1}{c}{Gap} & \\
\hline
FeGa$_{3}$ &    &   &   &      \\
LDA &  0.00/0.00  &0.00/0.00  & 0 &0.52    \\
LDA+$U$ &  1.07/-1.07  &1.07/-1.07  & 0 &0.47 \\
mBJ &  0.00/0.00  &0.00/0.00  & 0 &0.57     \\
\hline
\hline
FeGa$_{2.75}$Zn$_{0.25}$ &    &   & &        \\
LDA &  0.40/0.40  &0.00/0.00  & 0.8 &0     \\ 
mBJ &  1.00/1.00  &-0.50/-0.50  & 1.0 &0.04     \\
\hline
FeGa$_{2.5}$Zn$_{0.5}$ &    &   & &       \\
LDA &  0.75/0.75 &0.07/0.34 & 1.9 & 0  \\ 
mBJ &  1.29/1.29 &0.77/-0.95 & 2.0 & 0.14 \\

\hline
\hline
FeGa$_{2.75}$Ge$_{0.25}$ &    &   &&        \\
LDA &  0.22/0.22 &0.23/0.23 & 0.9& 0  \\
mBJ &  0.42/0.42 &0.20/0.20 & 1.0 &0 \\
\hline
FeGa$_{2.5}$Ge$_{0.5}$ &    &   &  &     \\
LDA &  0.43/0.36  & 0.40/0.40  & 1.7 &0     \\
\hline
\hline
 &    M Mn,Co/Fe2  &  M Fe3/Fe4 & M$_{tot}$  & Gap    \\
\hline
Fe$_{0.75}$Mn$_{0.25}$Ga$_{3}$&   & &  &      \\
LDA &  0.76/0.26  & 0/0  & 1.0 &0     \\
LDA+$U$ &  2.10/-0.86  &0.94/-1.16 & 1.0 &0 \\
mBJ &  1.28/0.38  &-0.30/-0.30  & 1.0 & 0     \\
\hline
\hline
Fe$_{0.75}$Co$_{0.25}$Ga$_{3}$&   & &  &       \\
LDA &  0.17/0.34  & 0.20/0.20  & 0.9 &0     \\
LDA+$U$ &  0.14/1.09  &-1.11/1.03 & 1.0& 0 \\
mBJ &  0.17/0.50 & 0.23/0.23  & 1.0 & 0     \\

\label{table1}
\end{tabular}
\end{ruledtabular}
\end{table*}

\section{Stoichiometric FeGa$_3$}

\begin{figure}[!ht]
\includegraphics[width=\columnwidth,draft=false]{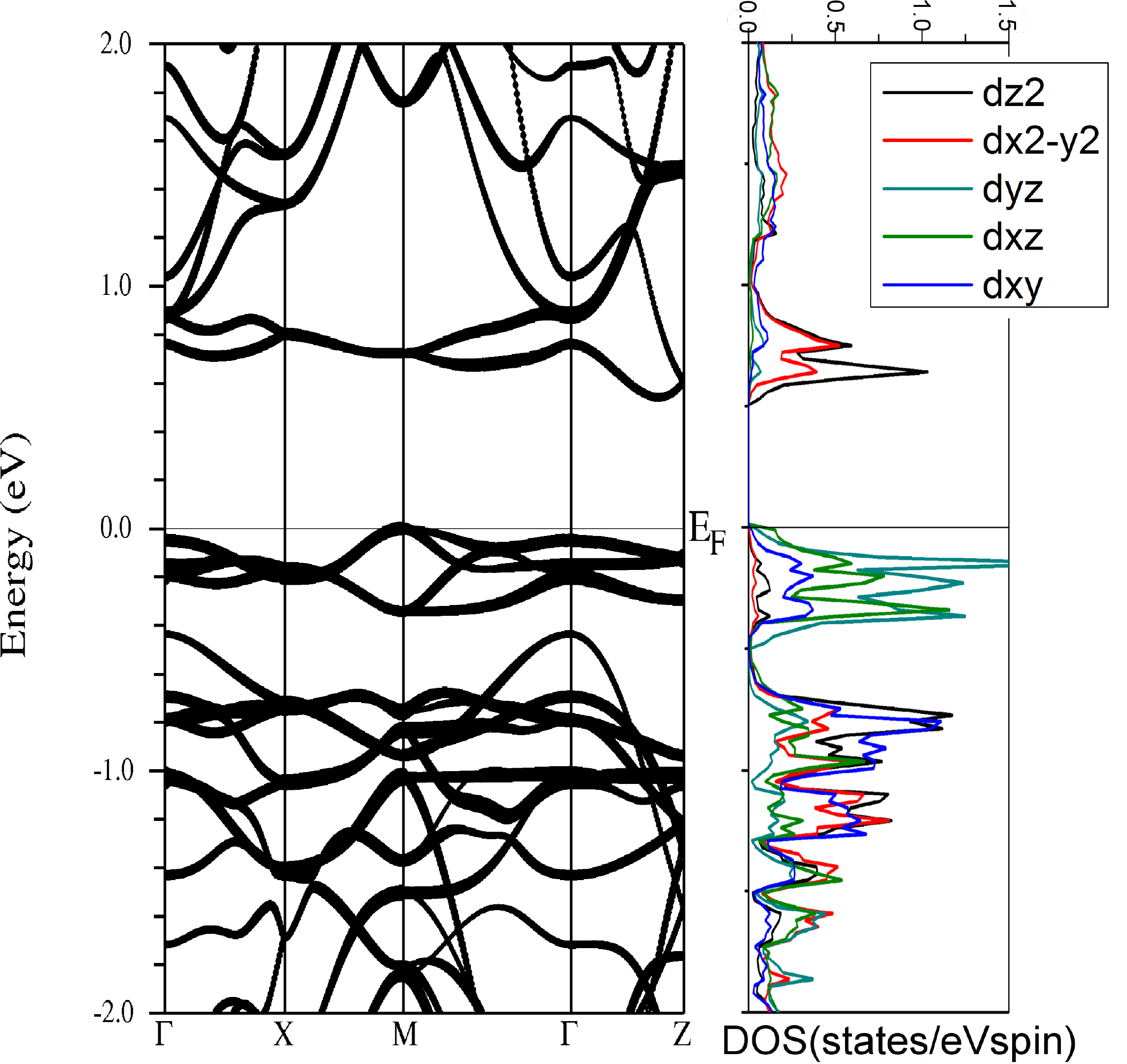}
\caption{(Color online)  Left panel: band structure with band character plot (Fe highlighted) of non-magnetic FeGa$_3$ 
within LDA, illustrating the
predicted band gap that is in agreement with experiment. Right panel: Orbital resolved DOS for Fe orbitals, in local 
coordinates (see text and Fig. \ref{struct}) for nonmagnetic FeGa$_{3}$ and  obtained within LDA. The Fermi level is at zero.}
\label{dos_lda}
\end{figure}

\textit{LDA.} Both ferromagnetic and antiferromagnetic orientations within the Fe$_2$ dimers have been studied.
 Within LDA (GGA) only a nonmagnetic state can be stabilized (see Table \ref{table1}). This result is consistent with the observed 
 diamagnetism in pure FeGa$_3$ and with previous electronic structure 
 calculations.\cite{PhysRevB.88.064422,PhysRevB.82.155202}  
In Fig. \ref{dos_lda}  the band structure with band character plot and DOS of FeGa$_3$ within LDA are shown. The 
electronic state is insulating with a band gap of 0.52 eV. 

Flat bands both below and above the gap give rise to a large DOS into which electrons or holes will be doped. 
Most of the contribution to states around the Fermi level comes from narrow Fe-$d$ bands with 
little Ga-$p$ contribution. For Fe $3d$ states we use a local coordinate system with the local $z$
axis oriented along the dimer axis (as shown in Fig. \ref{struct}).
Using this local coordinate system the DOS looks as shown in the right panel of 
Fig. \ref{dos_lda} with seemingly pseudo-cubic $d_{xy}$, $d_{xz}$, $d_{yz}$ occupation.
Below the gap,
occupied $d_{xz}, d_{yz}$ bands form a narrow, 0.35 eV wide four-band/spin complex.  
Above the gap, $d_{z^2}$ and $d_{x^2-y^2}$ 
bands form a 0.4 eV wide two-band/spin complex.
This different orbital character had been pointed out earlier
by Singh\cite{PhysRevB.82.155202} and by Yin and Pickett.\cite{PhysRevB.88.064422}

\textit{LDA+U.} For the undoped compound, when an onsite Coulomb repulsion is included, a magnetic state can only be obtained for an AFM ordering within the dimers at low $U$ (the magnetic moments vanish when FM order is set). The magnitude of the magnetic moment increases gradually as the $U$ value does (from 0.5$\mu_B$ for $U=$ 2 eV, to 1.07$\mu_B$ for $U=$ 3eV, and 1.63$\mu_B$ for $U=$ 4 eV). The derived magnetic moments are in agreement with those obtained by Yin \textit{et al.}\cite{PhysRevB.82.155202} An insulating state is retained for $U$ in the 2-4 eV range, but surprisingly the band gap closes as $U$  increases. The antiferromagnetic spin singlet scenario derived from LDA+$U$ calculations at $U$= 3 eV gives the same band gap as the LDA result ( with the DOSs differing only
in some specifics) and is consistent with transport and thermodynamic experiments (see Table \ref{table1}).

\subsection{Analysis of band character}
In halides, oxides, and some chalcogenides of iron, it is possible and very useful to identify
the charge state (also known as formal valence) of Fe. This underlying picture provides substantial
guidelines on the character of excitations that are likely to dominate the low energy behavior of
the system. In some unusual (semi)metallic transition metal compound, viz. CoSb$_3$,\cite{cosb3}
such identification proves to be difficult, with unconventional pictures arising. We preface our
study of doping of FeGa$_3$ with information relating to the Fe charge state (also called formal valence).
We remind that the formal charge state often has only a very indirect connection to the physical
charge density of atoms in the compound.\cite{quan_co}  

The integrated partial DOS, obtained from projections of the Bloch states, 
provides a guideline for the $3d$ occupation of Fe. Supposing that true
$3d$ character falls off above 1.5 eV, (i.e. higher lying $d$ character reflecting tails of Ga atoms
extending into the Fe sphere), the $d_{xy}$, $d_{xz}$, $d_{yz}$ orbitals are fully occupied, 6 e$^-$/Fe.
The other two orbitals, local $d_{z^2}$ and $d_{x^2-y^2}$ (which are very distinct orientationally
and probably chemically), are
75\% occupied, thus providing 3 $e^-$/Fe. The inferred occupation is then a surprising Fe$^{-1}$: $4d^9$.
This characterization would be (Fe$_2$)$^{-2}$ (Ga$_6$)$^{+2}$. Although this characterization is not outrageous --
Ga is quite electropositive and easily donates electrons to neighboring electronegative atoms, 
this picture should be given further scrutiny.

\subsection{Fe$-$Fe dimer from Wannier function perspective}

\begin{figure}[!ht]
    \includegraphics[width=\columnwidth]{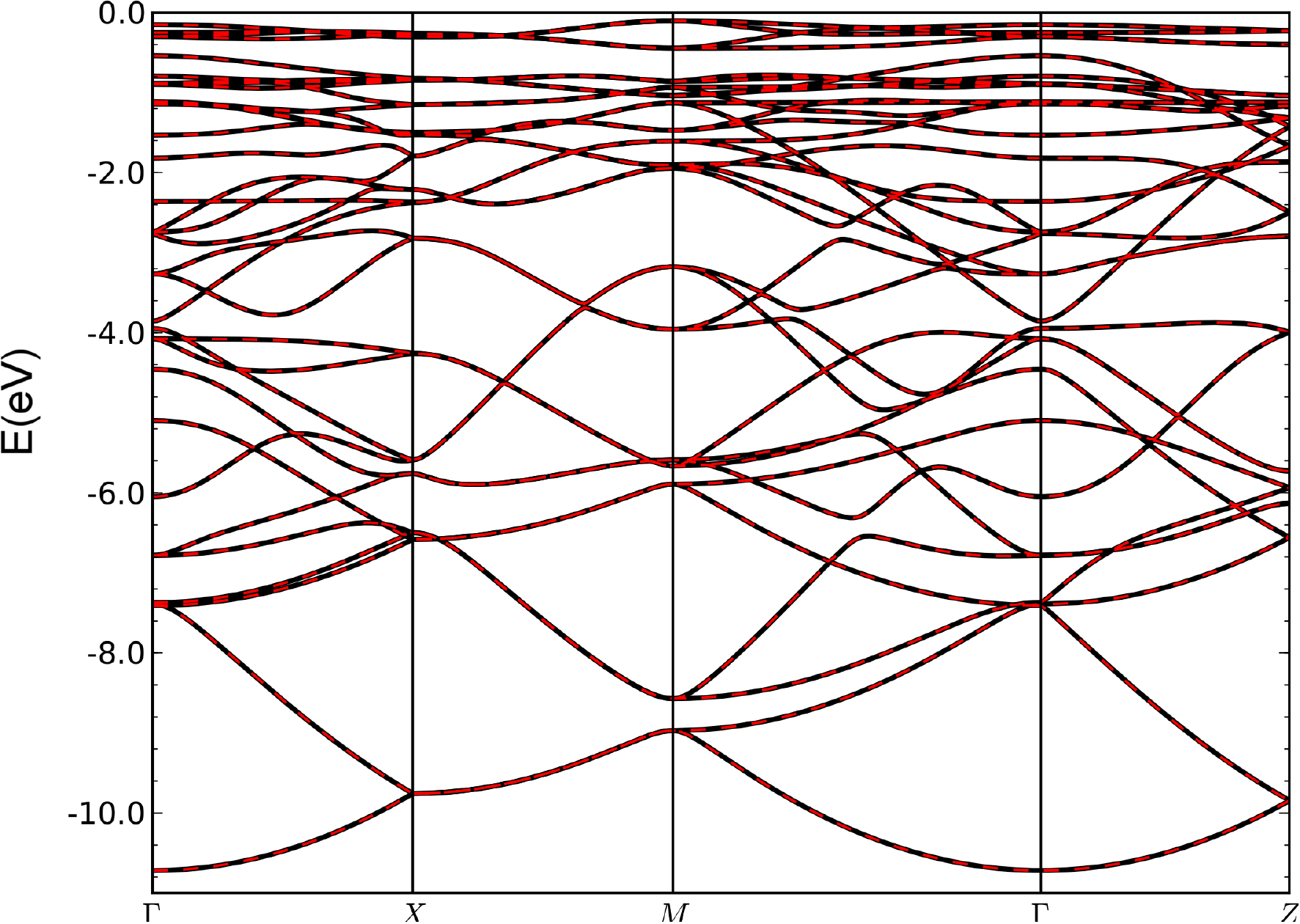}
    \caption{Comparison between the band structure obtained
        from the DFT calculation (in black) and the Wannier function interpolation (in red). The
        $k$-mesh used in Wannier function calculation does not  include
        $k$-points along some symmetry lines. Therefore, that the
        interpolated band structure from Wannier functions agrees
    with the Bloch bands provides a measure of the precision of the Wannier function
        transformation.}
\label{bs_wannier}
\end{figure}

\begin{figure}[!ht]

\includegraphics[width=\columnwidth]{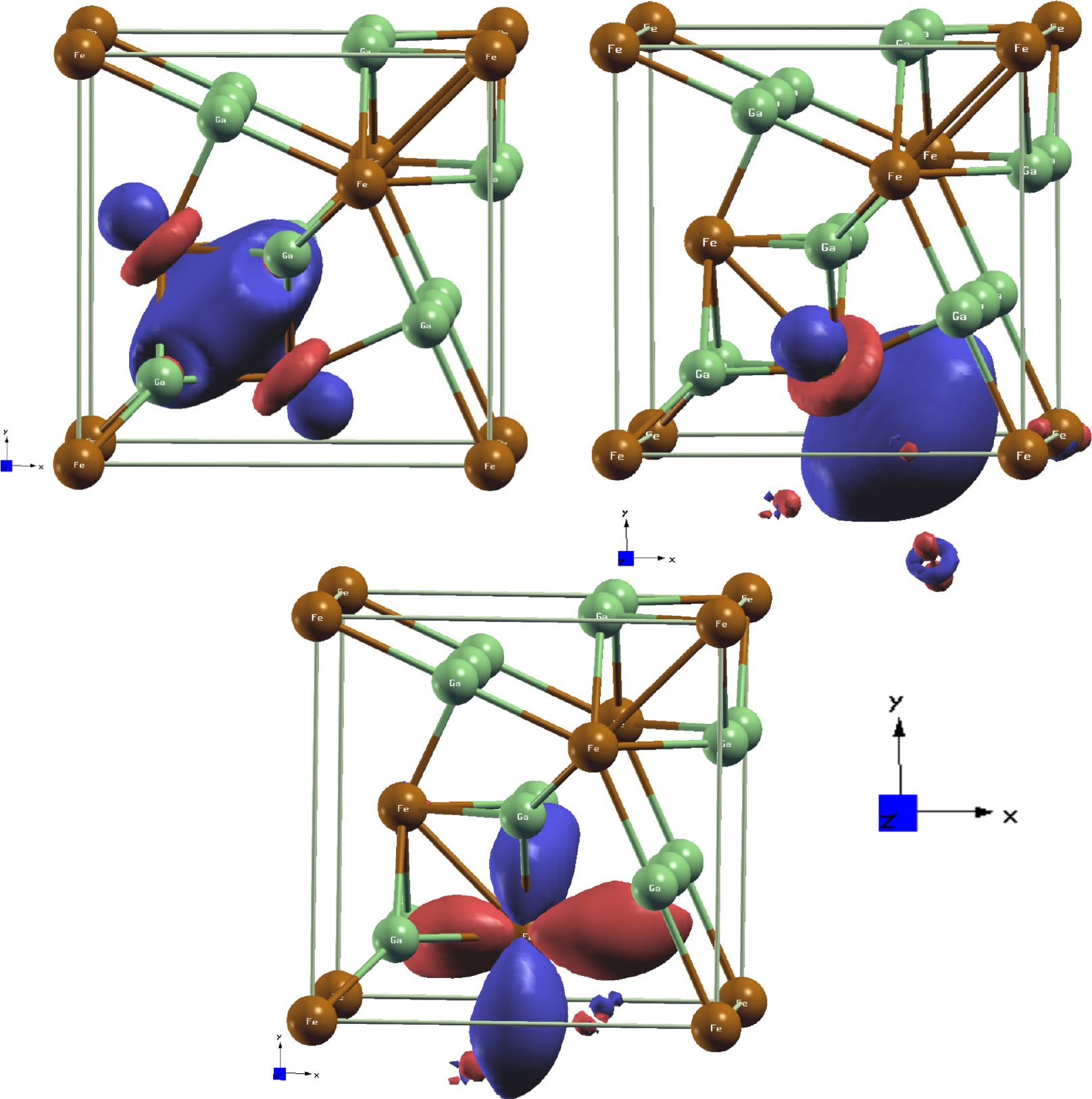}
\caption{Top left panel: Wannier function representing the Fe-Fe bonding orbital (for one Fe$_2$ dimer) linked to the bonding 
d$_{z^2}$ and d$_{x^2-y^2}$ orbitals occupied below the set of d$_{xz}$, d$_{yz}$, d$_{xy}$ orbitals (in local coordinate 
system). Top right panel: example of Wannier function shared between Fe-Ga atoms. The red loop lies on an Fe site. 
Bottom panel: example of $t_{2g}$-like Wannier function centered on an Fe site. Fe atoms in brown, Ga atoms in green. Isosurfaces of different color correspond to opposite signs of the WF}
\label{wf}
\end{figure}

There are, incontrovertibly, 17 occupied bands per spin channel per Fe$_2$ 
dimer cluster. This situation, and especially the gap obtained even in the weakly correlated treatment, strongly
suggests covalent (or metallic) bonding rather than ionic bonding as in many Fe
insulators. Since the starting procedure is to assign orbitals localized at specific sites in the
initial projection to obtain Wannier functions (WFs),\cite{wannier90, wien2wannier} our choice was to have (per spin) 4 electrons per Fe, two electrons
for six of the Ga atoms and one electron for the other six Ga atoms. Often in
complex structured materials these initial
associations do not persist, and that was the case here. The agreement between the occupied band structure obtained from Wannier function interpolation and that derived from the DFT calculation is excellent (see Fig. \ref{bs_wannier}), indicating a
faithful (though not unique) transformation to WFs. Four different initial projections resulted in four sets of WFs with the 
same total spread (mean localization).
A persistent feature throughout the sets was that, of the 34 WFs/spin,
precisely two are centered in the middle of dimers, i.e. one per
dimer as can be seen in the top left panel of Fig. \ref{wf}, or 1/2 electron per Fe. Most of the
 other 32 Wannier functions/spin resemble each other: lopsided objects
with much of the occupation assignable to the Fe site 
 (top right panel of Fig.\ref{wf}). In some cases 1-3 $t_{2g}$-like WFs 
centered on Fe result, and these have the lowest 
 spread of any of the WFs (lower panel of Fig. \ref{wf}). 

The ``dimer bonding WF'' introduces a peculiar and, to our knowledge, unprecedented
local orbital picture of an unpolarized metal-atom-based insulator, and is 
related directly to the dimer structure in the cell. It corresponds to a
one-center two-electron bond between the two Fe atoms, and is directly analogous
to the H$_2$ molecule (where the WF density is equivalent to the total density
per spin since there is only one orbital per spin). 
Since the character of the density just below the gap is antibonding, 
as is clear from Fig. 5, and the Fe $3d$ shell is more than half filled, this 
bonding dimer WF must arise primarily from deeper, bonding Fe states.  

We conclude therefore that no simple Fe charge state picture works for FeGa$_3$. The
strong separation between occupied pseudocubic ``$t_{2g}$'' orbitals and empty ``$e_g$''
orbitals is a distinguishing feature of the
Fe$_2$ dimer, as is strong mixing with the Ga $s-p$ orbitals.

\subsection{Charge density of flat near-gap bands}

\begin{figure}
\includegraphics[width=0.48\columnwidth,draft=false]{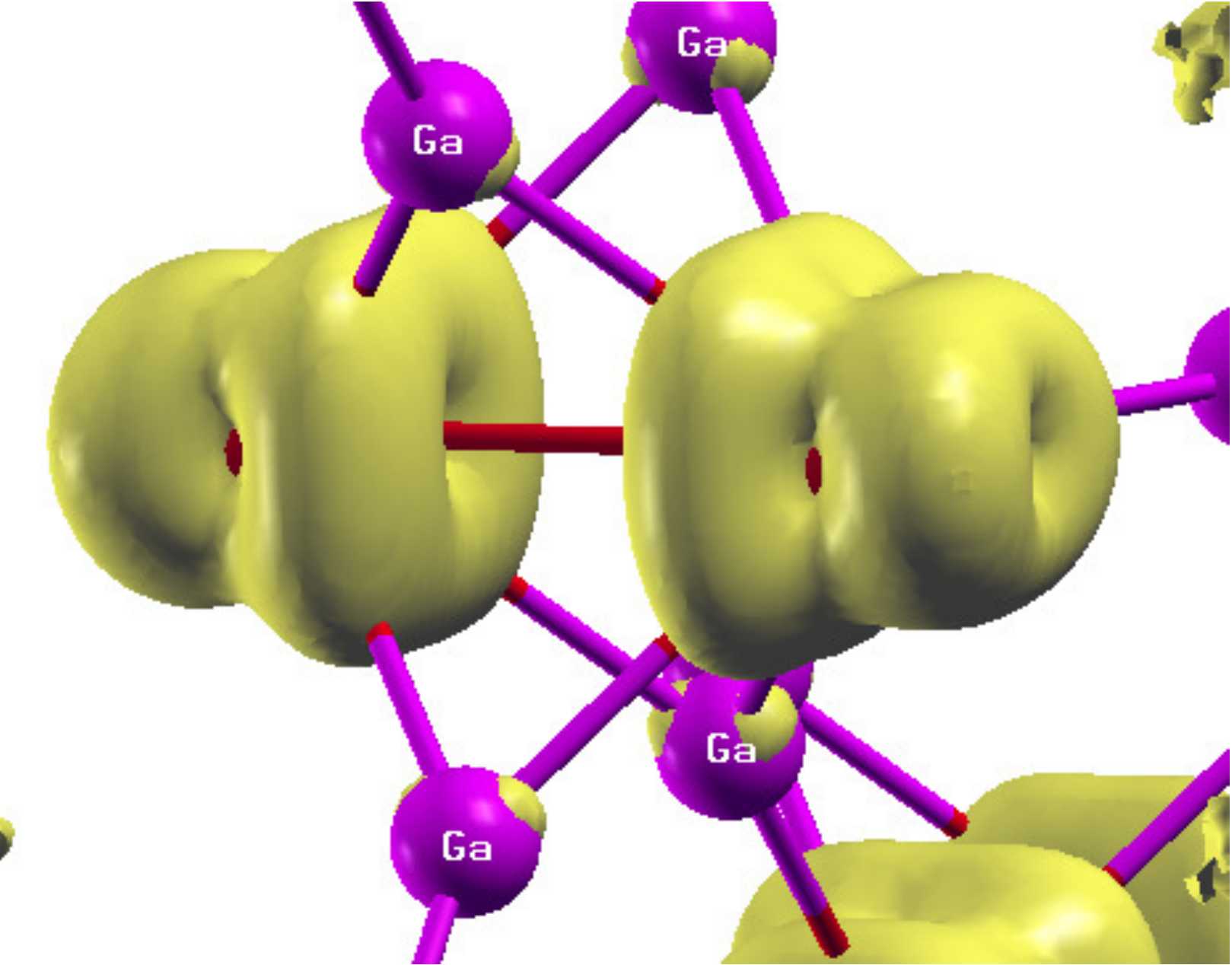}
\includegraphics[width=0.48\columnwidth,draft=false]{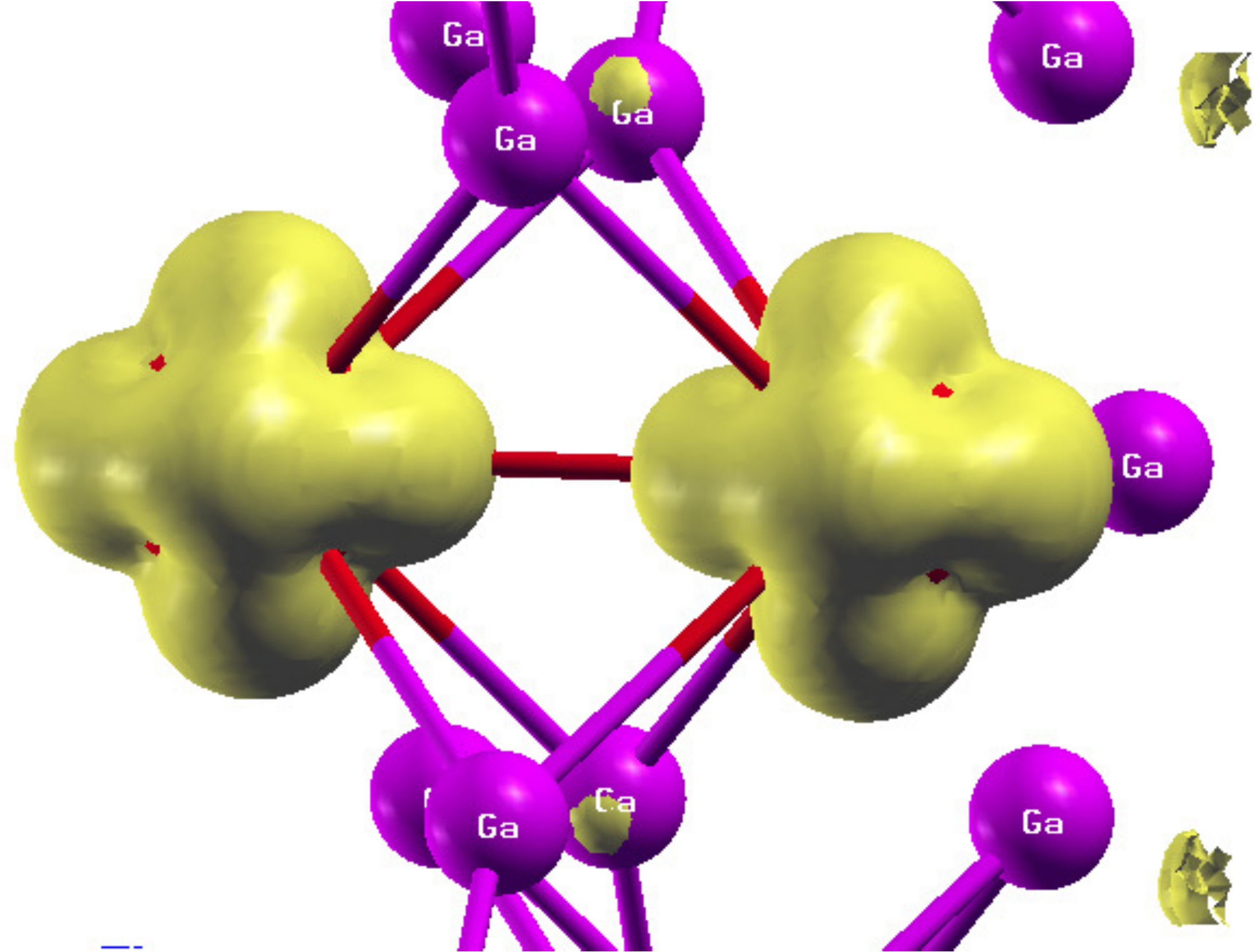}\\
\caption{\label{fignew}Isosurface plots of states in FeGa$_3$. Left panel: charge density obtained using XCRYSDEN
\cite{xcrysden} corresponding to the states at the bottom of the gap in FeGa$_3$ (energy range -0.3-0 eV). Right panel: charge density of the states 
 above the gap in FeGa$_3$ (energy range 0.6-0.9 eV). The strong directionality and the antibonding or nonbonding character of the charge 
along the Fe-Fe bond is evident. While the charge remains localized around Fe atoms, the differing characters of the 
contributing states is obvious.}
\label{rho}
\end{figure}

Figure \ref{rho} shows 3D isosurfaces of the charge density for the narrow  Fe $3d$-dominated bands
below and above the gap.  Consistent with the discussion provided above, 
the occupied states are a mixture of pseudocubic $d_{xz}$ and 
$d_{yz}$ charge ($z$ along the dimer axis), 
with smaller contribution from $d_{xy}$, and is roughly circular in cross section
around the dimer axis. 
The unoccupied states are a combination of pseudocubic $e_g$ states: $d_{z^2}$ orbitals
directed along the dimer axis, and $d_{x^2-y^2}$ with lobes perpendicular to the dimer axis. This
density is decidedly non-circular in cross section. Both densities reflect antibonding character, or at least
nonbonding character, with no evidence of bonding charge.
There is negligible contribution to the density that is displayed from Ga sites. The very small
dispersion of the bands bordering the gap suggests rather localized molecular orbitals as the basic underlying
feature for near-gap states.

\section{Doped FeGa$_3$}

We have performed calculations in which both $4p$ (Ga) and $3d$ (Fe) substitutional doping mechanisms have
 been explored in the 12-atom cell described above: either one Ge or one Zn were substituted on Ga1 or Ga2 sites (doping level $y$= 0.25, 0.5, respectively) and Fe was substituted by either 
 one Co or one Mn (doping level $x$= 0.25).

Within LDA our results are consistent with those presented by  Singh within GGA.\cite{PhysRevB.88.064422} Due to the 
steep and large DOS on either side of the gap,  an itinerant FM state is obtained in all cases of low doping 
(Stoner mechanism for itinerant ferromagnetism). However, the magnetic state is different for the two types of directions 
of doping, as we will describe below.  

The scenario in which semiconducting FeGa$_3$ has singlet Fe$_2$ dimers (an antiferromagnetic state in a DFT-based calculation)  was proposed by Yin and Pickett  based on LDA+$U$ calculations.\cite{PhysRevB.82.155202}  As mentioned above, for the stoichiometric compound, LDA and LDA+$U$ ($U$= 3 eV) give very similar band gaps consistent with the experiment. When doping occurs, and especially when substitution is done on the Fe
site, effects of correlation are likely to become more evident. We also pursue this
goal in the present section. 

Table \ref{table1} shows the magnetic moments for each of the atoms in the metal dimers, the total magnetic moment in the cell, and the band gap for each of the cases studied in the corresponding magnetic ground state.
This table will guide much of the discussion that follows.

\subsection{Hole doping Fe$\rightarrow$Mn and Ga$\rightarrow$Zn
within LDA}

\begin{figure}
\includegraphics[width=\columnwidth,draft=false]{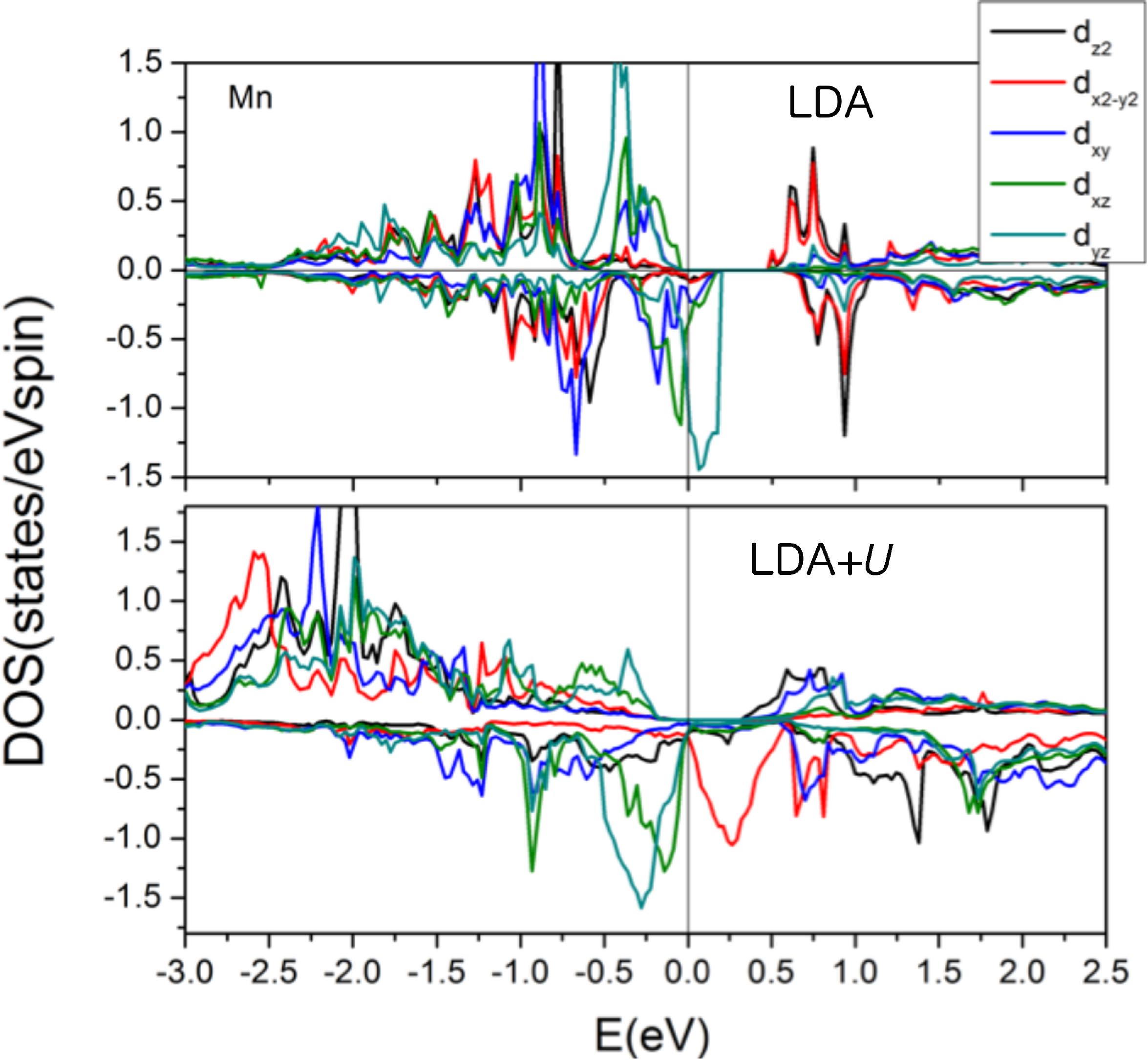}\\
\caption{Orbital resolved DOS of Mn atoms in local coordinates in the magnetic ground state of Fe$_{0.75}$Mn$_{0.25}$Ga$_3$  within LDA (upper panel) and LDA+$U$ ($U$= 3 eV) (lower panel). Hole doping shifts the Fermi level into the four-band complex with $d_{xy}$, $d_{xz}$, $d_{yz}$ character for the minority spin channel giving rise to a half metallic (HM) result. The onsite Coulomb repulsion causes orbital reoccupation with shifts in the band structure of 1 eV, and introduces $e_g$-like states in the gap for the minority spin channel. Positive/negative values of the DOS correspond to the spin-up/spin-dn channel. The Fermi level is set at zero.}
\label{bs_femn}
\end{figure}

Hole doping has been explored on the Fe site (replacing one Fe by Mn) and by  substitution on the Ga site by Zn: Ga1 substitution corresponds to a doping level $y$= 0.25, and the Zn atom has a pair of Fe atoms at a distance of 2.36 \AA. Ga2 substitution corresponds to a doping level $y$= 0.5, and the substituted Zn has also a pair of Fe atoms at a distance of 2.50 \AA ~and one of the Fe of the other dimer at 2.39 \AA ~(see Fig. \ref{struct}).

In both cases, only an itinerant FM state can be obtained within LDA. A half metallic (HM) FM state is obtained for Fe$_{0.75}$Mn$_{0.25}$Ga$_3$ with magnetic moments of 0.76 $\mu_B$ on the Mn atom, 0.26 $\mu_B$ on the paired Fe, and the other Fe$_2$ dimer remaining non magnetic (see Table \ref{table1}). The corresponding density of states is shown in the upper panel of Fig. \ref{bs_femn}. It exhibits a simple exchange splitting plus rigid-band behavior characteristic of an itinerant FM. The single hole difference leaves E$_F$ lying within the four $3d$ bands of the metallic minority spin channel, which are very strongly Mn character -- the hole remains on the Mn atom. Magnetism arises from the $t_{2g}$ orbitals, with larger contribution from the d$_{yz}$ orbital at the Fermi level. The exchange splitting is about 0.25 eV. 

For FeGa$_{2.5/2.75}$Zn$_{0.5/0.25}$ an itinerant FM state emerges with magnetism arising also from the $d_{yz}$ orbital. The hole is localized on the Fe atoms closer to the substituted Zn (see Table \ref{table1} always showing higher magnetic moments on the Fe atoms closer to the Zn, with the other Fe atoms having a significantly lower magnetic moment or even being nonmagnetic). 

Experimental data on Mn- and Zn-doped FeGa$_3$ differ from the LDA-based predictions of emergence of a conducting FM state. $p$-type doping shifts the Fermi level into the valence band and gives rise to metallic behavior, whereas transport measurements for hole doped FeGa$_3$ show no insulator-metal transition, but instead semiconducting behavior, not consistent with itinerant behavior.  

\subsection{Hole doping Fe$\rightarrow$Mn 
within LDA+$U$}
 
For hole (Mn) doping on the Fe site, for the lower $U$ value of 2 eV, only FM ordering within the pairs can be obtained.  A strong moment of 1.43$\mu_B$ on Mn 
(S=$\frac{3}{2}$) is obtained. The resulting electronic structure has the same features as that obtained within LDA: a HM FM state. Thus at $U$=2 eV correlation effects do not appear to be significant.

For $U \geq$ 3 eV the picture changes qualitatively. Not only can states with the magnetic moments being aligned and antialigned be obtained, but the AFM alignment becomes energetically favored.  The lower panel of Fig. \ref{bs_femn} 
shows the LDA+$U$ ($U$= 3 eV) orbital resolved density of 
states of the Mn atom in Fe$_{0.75}$Mn$_{0.25}$Ga$_3$  for the energetically favored AFM alignment within the dimers. The magnetic moments are strengthened with $U$ (see Table \ref{table1}). While HM character remains,
the onsite Coulomb repulsion shifts certain $d$ bands by more than 1 eV 
giving rise to distinctive changes in configuration. A band with $d_{x^2-y^2}$ character 
appears within the gap in the minority spin channel, in contrast to the LDA 
case. The complete
$d_{x^2-y^2}$ spectral density within the gap strongly implies that for an
isolated Mn dopant ({\it i.e.} the low doping limit) an unoccupied $d_{x^2-y^2}$
gap state will appear, perhaps near mid-gap, when strong correlations are considered.

The band shifts are even more noticeable for higher $U$ values, 
with the majority gap closing accompanied by an 
increase in the magnetic moments of Mn and Fe that reach values of 3.00$\mu_B$ 
and 2.05$\mu_B$, respectively, and a widening of the bands around the Fermi level.

\subsection{Electron doping Fe$\rightarrow$Co and Ga$\rightarrow$Ge
within LDA}

\begin{figure}[!ht]
\includegraphics[width=0.95\columnwidth]{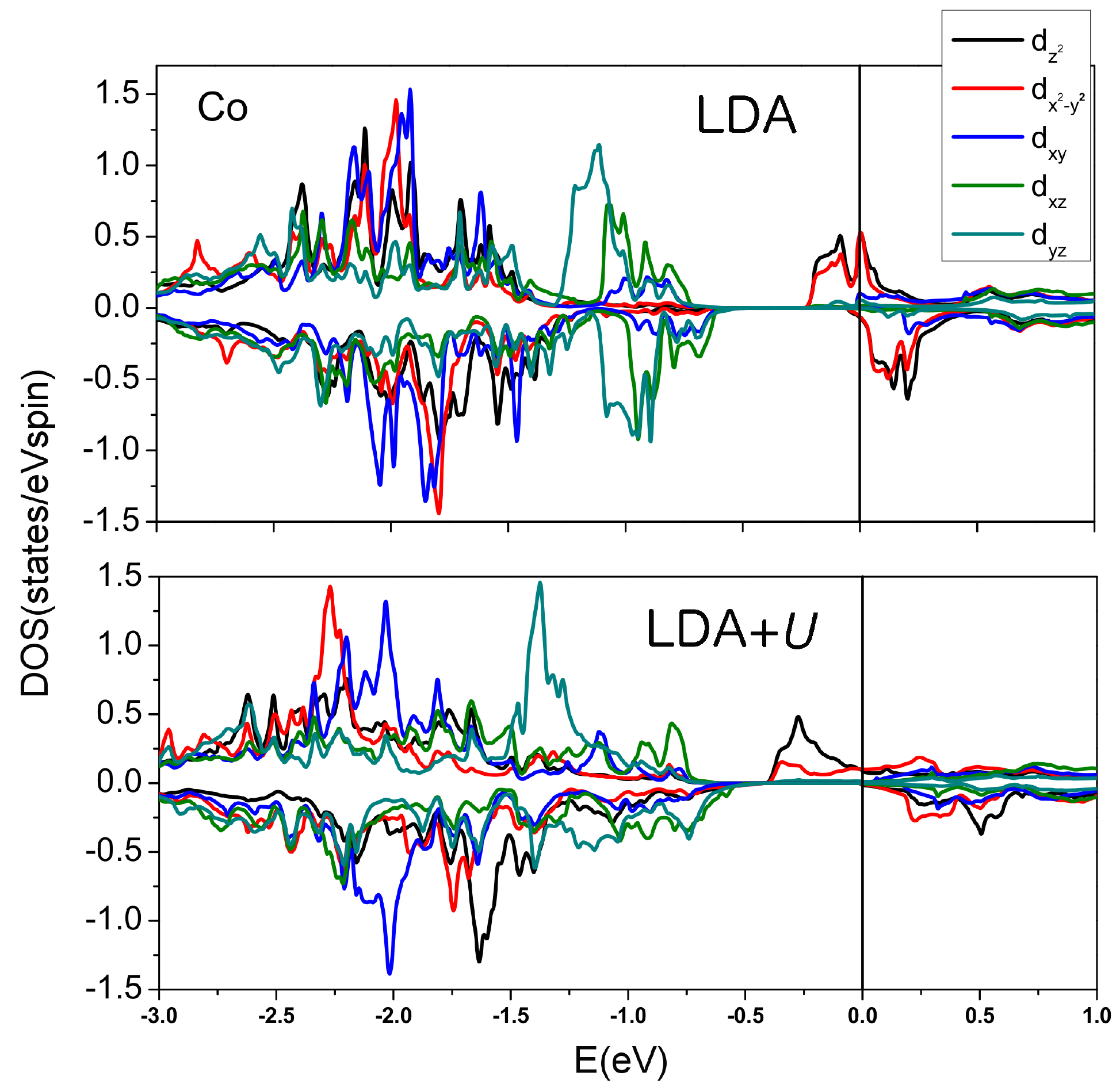}
\caption{Orbital resolved DOS of Co atoms in local coordinates in the magnetic ground state of Fe$_{0.75}$Co$_{0.25}$Ga$_3$  within LDA (upper panel) and LDA+$U$ (lower panel). Electron doping primarily drives the Fermi level into the conduction band giving rise to a half metallic state upon doping. Positive/negative values of the DOS correspond to the spin-up/spin-dn channel. The Fermi level is set at zero.}\label{dos_feco}
\end{figure}

\begin{figure}[!ht]
\includegraphics[width=\columnwidth]{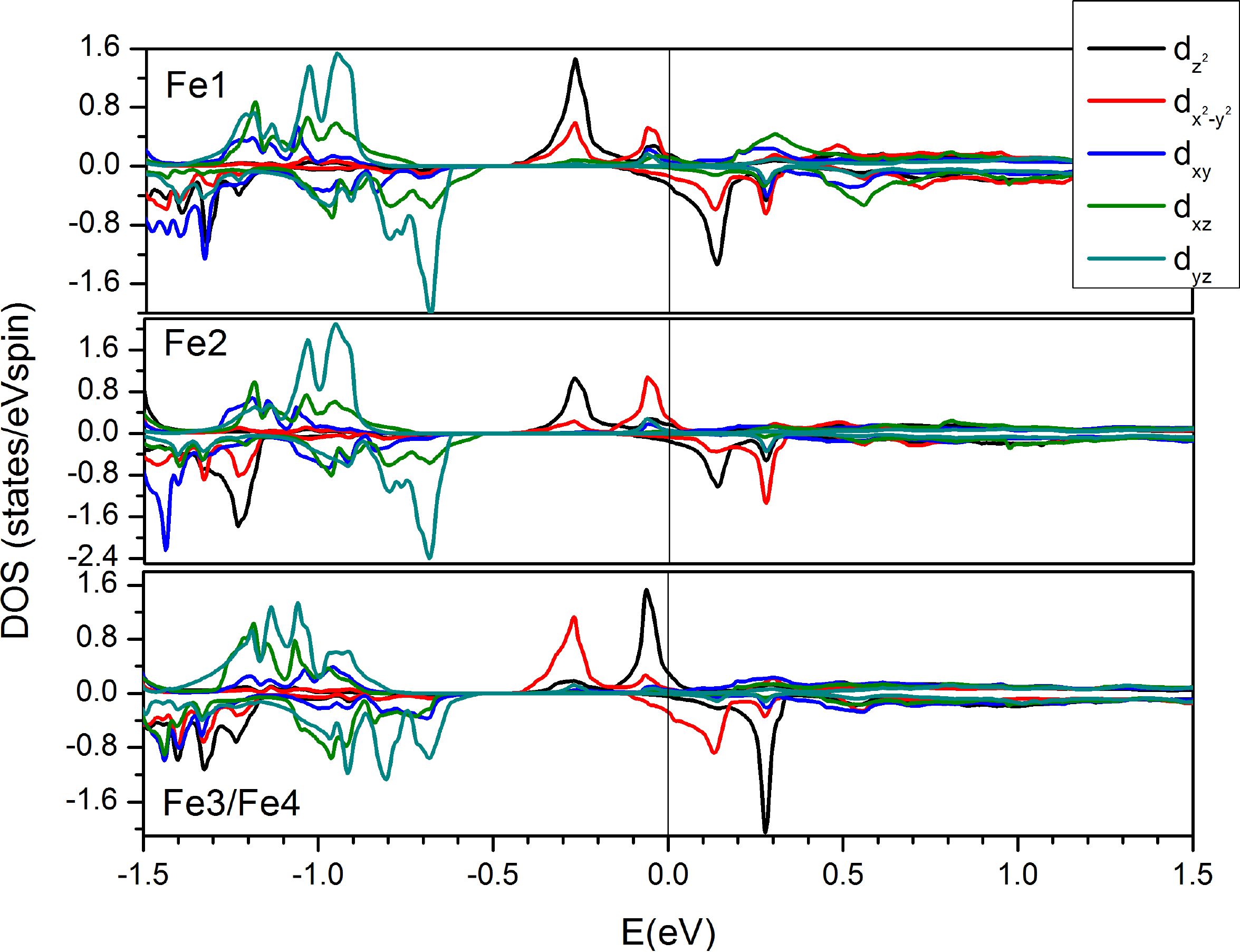}
\caption{Orbital resolved DOS of the non-equivalent Fe atoms in local coordinates in the magnetic ground state of FeGa$_{2.5}$Ge$_{0.5}$ obtained within LDA. The upper two panels show the DOS of Fe1 and Fe2 with partially occupied $d_{z^2}$ and $d_{x^2-y^2}$ orbitals. The lower panel shows the DOS at Fe3/Fe4 sites. Electron doping through Ge substitution pushes the Fermi level into the conduction band for both spin channels and induces magnetic moments on Fe supported by exchange splitting. Positive/negative values of the DOS correspond to the spin-up/spin-dn channel. The Fermi level is set at zero. }
\label{dos_ge}
\end{figure}

Electron doping at the Fe site (by Co) and at the Ga site (by Ge) also gives rise to an itinerant FM state within LDA. In this case, magnetism arises from the $e_g$-like orbitals and the total moment of 1 $\mu_B$ is spread rather equally across all transition metal ions -- the electron is less localized (see Table \ref{table1}). There are only small differences in the DOS for electron doping on the Fe or Ga site: low level electron doping with Co for Fe gives a simple rigid band shift downward of the majority spin channel; when doping with Ge different splittings appear for $d_{z^2}$, $d_{x^2-y^2}$ orbitals due to the change in local environment with respect to the Co-doped compound (compare top panel of Fig. \ref{dos_feco} for Fe$_{0.75}$Co$_{0.25}$Ga$_3$ with Fig. \ref{dos_ge} for FeGa$_{0.5}$Ge$_{0.5}$).

From experiments, Co-doped FeGa$_3$ behaves differently from the Ge-doped counterpart. The Ge-doped compound is ferromagnetic with a positive Curie-Weiss temperature whereas experiments for Co-doping show an AFM ordering at low T.   DFT-based calculations do not account for that difference, always favoring a ferromagnetic alignment of moments. The main difference with respect to hole doping is the more delocalized picture with the moments being equally distributed among all the metal atoms, in contrast to hole doped FeGa$_3$ where there is a greater localization of the magnetic moments.

\subsection{Electron doping Fe$\rightarrow$Co 
within LDA+$U$}

For the lower $U$ value of 2 eV correlation effects are not significant in electron doped FeGa$_3$: a  HM FM state is derived. 
At $U=$ 3 eV a ground state with antialigned moments is obtained. The lower panel of Fig. \ref{dos_feco} shows the orbital resolved DOS of the Co atom in Fe$_{0.75}$Co$_{0.25}$Ga$_3$ for the LDA+$U$ ($U$= 3 eV) state. The picture upon 
electron doping appears somewhat simpler than for hole doping: the various
orbital projected densities of 
states can be visualized as being connected adiabatically to their $U$=0 
position (upper panel) with a HM state persisting. The outstanding difference is that the occupied
$d_{z^2}$+$d_{x^2-y^2}$ weight above the gap converts to predominantly
$d_{z^2}$. This change corresponds to Co introducing an occupied $d_{z^2}$
majority gap state into the gap at low doping. The Co ion remains low-spin, whereas the 
magnetic moment on Fe strengthens upon introducing $U$(see Table \ref{table1}). 

For $U$ $\geq$ 4 eV noticeable shifts in the $d$-bands take place, with the Fe moment increasing up to 2.05 $\mu_B$ for  $U$= 5 eV while Co remains in its low spin state, with induced moment of 
0.2$\mu_B$ irrespective of the $U$ value over the range we studied.

\section{Discussion}

As mentioned above, there is no unambiguous picture about the presence/absence of magnetism and/or correlations in FeGa$_3$. In agreement with previous works, the observed gap for undoped FeGa$_3$ (formed between t$_{2g}$ pseudocubic states at the top of the valence band and e$_g$ states at the bottom of the conduction band) can be reproduced without the inclusion of magnetism and correlations. 
This analysis is consistent with susceptibility measurements that indicate a diamagnetic state below RT\cite{gap_1} that is also supported by the absence of local moments in Fe Moessbauer spectra. \cite{mossbauer}

The antiferromagnetic spin singlet scenario derived from LDA+$U$ calculations ($U$= 3 eV) is consistent with transport and thermodynamic experiments, combined with neutron powder diffraction data, interpreted as showing that FeGa$_3$ is a correlated band insulator with a complex antiferromagnetic ordering.\cite{PhysRevB.89.195102} Photoemission experiments agree with the correlated picture since the observed band dispersions are well reproduced by calculations within the LDA+$U$ scheme ($U$= 3 eV).\cite{arita} The presence of magnetism in undoped FeGa$_3$ is also consistent with muon spin rotation studies that show a spin polaron band that requires the existence of Fe moments.\cite{muon}

Electron or hole doping described without including correlations gives rise to an itinerant FM state (Stoner magnetism) without pre-formed moments in the undoped compound. This type of result is consistent with experiments that show that FeGa$_{3-y}$Ge$_y$ is a weak itinerant ferromagnet, but are difficult to reconcile with the role of strong correlations in the FM instability found beyond y$_c$= 0.13. \cite{PhysRevB.86.144421}

This calculated metallic FM state also contrasts with experiments in Co-doped FeGa$_3$ that show itinerant antiferromagnetic behavior and suggest the existence of in-gap states at low doping.\cite{PhysRevB.89.104426} The correlated picture shows that Co substitution for Fe introduces a localized electronic gap state and gives rise to antialigned moments within the dimers, very different from the uncorrelated picture and supporting the experiments. 

Experimental data on Mn and Zn doping also differ from the LDA-based predictions of emergence of a HM FM state.  Gamza \textit{et al}\cite{PhysRevB.89.195102} showed that hole doping does not give rise to an insulator-to-metal transition. They found instead that both substitution of Zn onto the Ga site, or Mn on the Fe site introduces states into the semiconducting gap that remain localized, suggesting the formation of small magnetic polarons. Using neutron powder diffraction measurements and DMFT, they establish that the complex magnetic order for FeGa$_3$ above room temperature is almost unaffected by hole doping even though dynamical correlation effects become stronger. Some of these observations are again in better agreement with the correlated picture that suggests that Mn doping produces an in-gap hole state and an AFM alignment within the dimers (even though the calculated state within LDA+$U$ is HM).

\section{Summary}

To summarize, density functional theory methods have been applied to follow the
evolution of the magnetic and electronic properties of FeGa$_3$ with doping, where
experiment shows that complex behavior including magnetism and quantum critical
behavior arises. A specific interest here is to probe the difference for both
electron and hole doping when doping is done on the magnetic Fe (dimer) site or on the
surely uncorrelated Ga site. Doping on either site introduces gap
states or moves the chemical potential into bands on either side of the gap.

Using conventional DFT for weakly correlated materials,
the behavior upon substitution on the Fe site (disturbing the dimers) or on the Ga site (doping off the dimers) is similar: an itinerant ferromagnetic state emerges, as might be anticipated with this approach.  However, there
are clear differences between electron and hole doping: whereas electron doping
gives rise to a more itinerant and delocalized picture (with magnetic moments evenly
distributed for all the Fe atoms in the unit cell), upon hole doping there is a
higher degree of localization of the magnetic moments -- moments develop on the
dimer closest to the dopant atom. Evidently the response to
doping of states below the gap is quite different from that of states above
the gap.

When applying the correlated LDA+U method, hole doping again is quite different
from electron doping. Co always maintains its low spin state, while Mn is
strongly magnetic. Whether this difference reflects differences in the host states
above and below the gap, or is due more to differences between Mn and Co, remains
unclear. Strong correlation effects lead to inter-dimer effects: doping on one dimer
induces strong moments on the other (``undisturbed'') Fe$_2$ dimer.

Correlation effects  are needed to be able to reconcile theoretical
predictions with some of the rich behavior observed in doped FeGa$_3$.
Understanding the interplay between localized and itinerant magnetism (including in-gap
states at low doping) will be crucial to explain the peculiar puzzles posed by this
intermetallic compound.

\section{Acknowledgments}
We acknowledge helpful communication on this topic with D. J. Singh. 
A.S.B. was supported by DOE Grant DE-FG02-04ER46111, 
Y.Q. and W. E. P. were supported by NSF Grant DMR-1207622.

\section{Appendix}

\begin{figure}
\includegraphics[width=\columnwidth,draft=false]{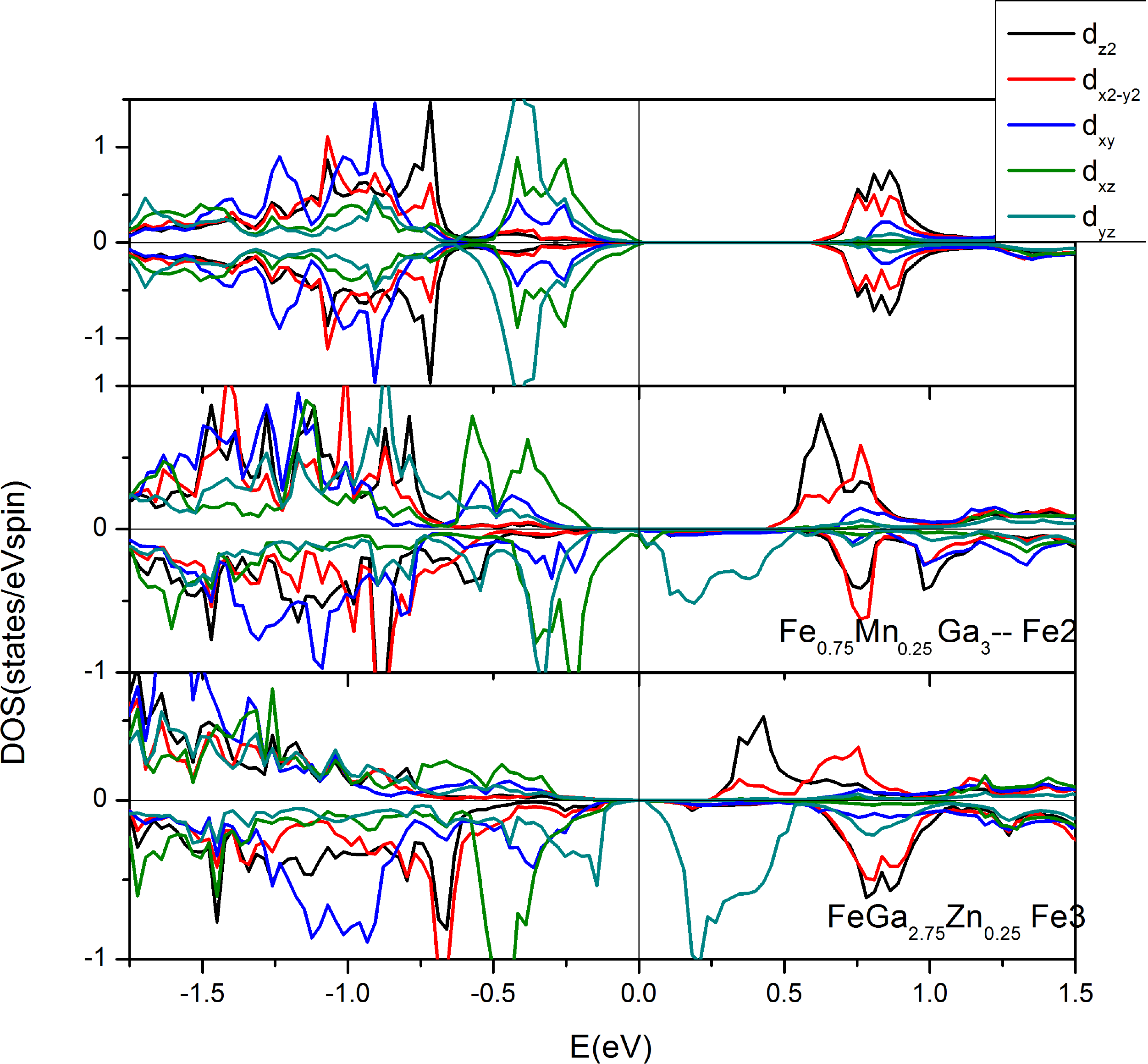}
\caption{\label{fignew} DOS in local coordinates for each Fe in the unit cell of FeGa$_3$ (top), Mn in Fe$_{0.75}$Mn$_{0.25}$Ga$_3$ (middle), and Fe in FeGa$_{2.75}$Zn$_{0.25}$  (lower panel) within mBJ. Half metallicity is broken in the doped cases (in contrast to the LDA predictions). Positive/negative values of the DOS correspond to the spin-up/spin-dn channel. The Fermi level is set at zero.}
\label{mbj}
\end{figure}

Gap closings are of particular interest in these studies of doping. To better
represent gap behavior, we have applied the LDA+ mBJ 
potential functional.
For stoichiometric FeGa$_3$ the band gap value obtained using mBJ is 
slightly increased with respect to that obtained within LDA. The DOS obtained within mBJ has the same features as that obtained within LDA (compare top panel of Fig. \ref{mbj} with Fig. \ref{dos_lda}).

When electron doped, the experimental and LDA-derived states are metallic and that type of state remains within mBJ. From Table \ref{table1} mBJ favors a more localized picture with respect to LDA or LDA+$U$. For Ge-doped FeGa$_3$ the magnetic moments are disproportionated within mBJ, being higher in the Fe pair (Fe1/Fe2) closer to the substituted Ge, unlike the LDA result where the moments were evenly distributed among all Fe atoms in the cell). 
For Co-doped FeGa$_3$, there is a slight increase in the magnetic moment of the Fe paired with the Co whose magnetic moment remains unchanged within mBJ, as do the magnetic moments of the Fe atoms in the other dimer.

When hole doped, the electronic structure obtained within mBJ is quite different from that derived within LDA and closer to the experimental results: the half metallicity is broken for both mechanisms of hole doping and antialigned magnetic moments within the dimers are favored.

For Fe$_{0.75}$Mn$_{0.25}$, the state resulting from mBJ consists of a FM ordering within both the Mn-Fe  and Fe-Fe pairs but with the moments having different sign for the two different dimers. The result is a zero-gap state versus the half metallic one derived using LDA. The corresponding DOS plot is shown in the central panel of Fig. \ref{mbj}. Clearly, mBJ shifts bands breaking the half metallicity, this is possible due to the shift of the d$_{yz}$-like orbital for the minority spin channel above the Fermi level for Mn and its paired Fe. Unlike the LDA result, the mBJ outcomes are consistent with experiments favoring a gap opening and an AFM alignment of the moments in the two different dimers for Mn-doped FeGa$_3$.

The response obtained for Zn-doped FeGa$_3$ within mBJ is also different from that of LDA. Again, the half metallicity is broken by mBJ and a semiconducting state arises (see bottom panel of Fig. \ref{mbj} for $x$= 0.25) with the concomitant increase in the magnetic moments that are antialigned as shown in Table \ref{table1}. 
For both doping levels the gap opening is again possible due to the splitting of the d$_{yz}$-like orbital for the minority spin channel of Fe atoms, being shifted above the Fermi level.

As commonly found in other applications, mBJ tends to increase the values of the magnetic moments (Table \ref{table1}).\cite{PhysRevB.83.195134} The shifts in the band structure and the increase in the magnetic moments with respect to LDA  are consistent with changes in the charge inside the atomic spheres: increased for Fe and accordingly reduced for Ga atoms.

\end{document}